\documentclass[11pt,twoside,a4paper,cmspaper,final,collab]{cms-tdr}
\def\svnVersion{dd08160}\def\svnDate{2026/08/27}\def\cmsCernNoTag{CERN-EP-2026-102}\def\cmsCernDate{\today}\def\cmsMessage{Published in Physical Review Letters as \href{http://dx.doi.org/10.1103/gdbv-wnpp}{\doi{10.1103/gdbv-wnpp}.}}
\begin{document}\cmsNoteHeader{SMP-24-014}

\ifthenelse{\boolean{cms@external}}{\providecommand{\cmsLeft}{upper\xspace}}{\providecommand{\cmsLeft}{left\xspace}}
\ifthenelse{\boolean{cms@external}}{\providecommand{\cmsRight}{lower\xspace}}{\providecommand{\cmsRight}{right\xspace}}

\newcommand{\dRlGoverPtG}{\ensuremath{\DR(\Pgg, \Pell)/(\pt^\PGg)^2}}

\newcommand{\phNP}{\ensuremath{\PGg_\text{NP}}}
\newcommand{\qqZZ}{\ensuremath{\PQq\PAQq\to\PZ\PZ}}
\newcommand{\qqZZll}{\ensuremath{\qqZZ \to 4\Pell}}
\newcommand{\qqZZnonpro}{\ensuremath{\qqZZ \phNP}}
\newcommand{\ggZZ}{\ensuremath{\Pg\Pg\to\PZ\PZ}}
\newcommand{\ggZZnonpro}{\ensuremath{\ggZZ \phNP}}
\newcommand{\ZZG}{\ensuremath{\PZ\PZ\PGg}}
\newcommand{\ZZGtollllG}{\ensuremath{\ZZG\to4\Pell\PGg}}
\newcommand{\pptoZZGtollllG}{\ensuremath{\Pp\Pp\to\ZZGtollllG}}
\newcommand{\pptollllG}{\ensuremath{\Pp\Pp\to4\Pell\PGg}}
\providecommand{\Pellpr}{{\HepParticle{\Pell}{}{\prime}}\Xspace}
\providecommand{\Pellprp}{{\HepParticle{\Pell}{}{\prime{+}}}\Xspace}
\providecommand{\Pellprm}{{\HepParticle{\Pell}{}{\prime{-}}}\Xspace}
\providecommand{\PZoptst}{{\HepParticle{\PZ}{}{(\ast)}}\Xspace}
\providecommand{\COMBINE}{\textsc{Combine}\xspace}

\newcommand{\midrule}{\noalign{\vspace{.4ex}\hrule height .05em\vspace{.65ex}}}

\newcommand{\makecell}[2][l]{{\renewcommand{\arraystretch}{1}\begin{tabular}{#1}#2\end{tabular}}}

\newlength\cmsTabSkip\setlength{\cmsTabSkip}{1ex}

\cmsNoteHeader{SMP-24-014}
\title{Evidence of \texorpdfstring{$\PZ\PZ\PGg$}{ZZgamma} production and observation of \texorpdfstring{$4\Pell\PGg$}{4lgamma} in proton-proton collisions at \texorpdfstring{$\sqrt{s} = 13\TeV$}{sqrt(s) = 13 TeV}}

\date{\today}

\abstract{Evidence of the production of two \PZ bosons and a photon in proton-proton collisions is reported for the first time in CMS. The analysis uses data collected by the CMS experiment between 2016 and 2018 at a center-of-mass energy of 13\TeV, corresponding to an integrated luminosity of 138\fbinv. Evidence for the process $\pptoZZGtollllG$ ($\Pell = \Pe$, $\PGm$), with an observed (expected) significance of 3.7\,(3.1) standard deviations in a fiducial region defined by $\pt^\PGg > 20\GeV$, $\abs{\eta^\PGg} < 2.4$, $\DR(\Pell,\PGg) > 0.5$, $m_{\PZ}$ between 60 and 120\GeV, and the invariant mass of either of the two \PZ bosons combined with the photon ($m_{\PZ\PGg}$) larger than 100\GeV, is reported. The measured (predicted) fiducial cross section is $60^{+27}_{-22}\unit{ab}$ ($47.56 \pm 0.04\unit{ab}$). Additionally, the inclusive production of $\pptollllG$ is studied by removing the $m_{\PZ\PGg}$ requirement to include final state radiation where one \PZ boson decays to $2\Pell\PGg$, yielding an observed (expected) significance of 5.0\,(4.2) standard deviations and a measured (predicted) fiducial cross section of $156^{+39}_{-35}\unit{ab}$ ($99.97 \pm 0.09\unit{ab}$).
}

\hypersetup{%
pdfauthor={CMS Collaboration},%
pdftitle={Evidence of ZZ gamma production and observation of 4 ell gamma in proton-proton collisions at sqrt(s) = 13 TeV},%
pdfsubject={CMS},%
pdfkeywords={CMS, electroweak interaction, W and Z bosons, photons, particle data analysis, triboson}%
}

\maketitle

Measurements of rare electroweak (EW) processes provide stringent tests of the standard model (SM) and may reveal physics beyond it.
The production of three EW bosons in a single proton-proton ($\Pp\Pp$) collision is
among the rarest processes accessible at the CERN LHC, with cross sections
of the order of tens of attobarns for the rarest channels.
Triboson processes include contributions from triple and quartic gauge couplings at tree level, as well as diagrams mediated by the Higgs boson.

The ATLAS~\cite{ATLAS-STDM-2021-09,ATLAS-STDM-2018-33} and CMS~\cite{CMS-SMP-19-013}
Collaborations have observed the production of two photons with a massive vector boson at 13\TeV.
Both collaborations have also reported observation of $\PW\PZ\PGg$~\cite{ATLAS-STDM-2019-17,CMS-SMP-22-018}
and of~$\PW\PW\PGg$~\cite{ATLAS-STDM-2019-28,CMS-SMP-22-006}.
Production of three massive EW bosons has been observed by both ATLAS~\cite{ATLAS-STDM-2017-22} and CMS~\cite{CMS-SMP-19-014}.
ATLAS has recently presented evidence for $\PZ\PZ\PGg$~\cite{ATLAS-STDM-2023-14}.
Comprehensive reviews of triboson measurements can be found in Refs.~\cite{CERN-EP-2024-075,CMS-SMP-23-004}.

\begin{figure*}[t]
  \centering
  \includegraphics[width=0.3\textwidth]{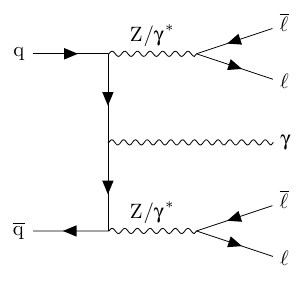}
  \includegraphics[width=0.3\textwidth]{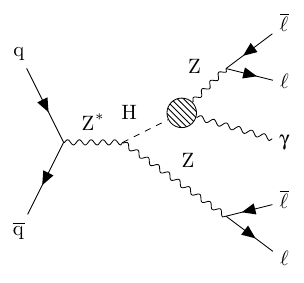}
  \includegraphics[width=0.3\textwidth]{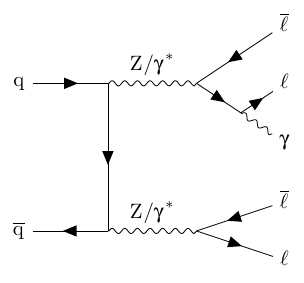}
  \caption{
    Representative tree-level standard model Feynman diagrams with four isolated leptons and a photon in the final state.
    From left to right,
    photon emitted from a quark line,
    associated $\PZ\PH$ production,
    and final-state radiation.
  }
  \label{fig:ppTo4LG}
\end{figure*}

In the EW production of four leptons (electrons or muons) and a photon,
$\Pellp\Pellm\Pellprp\Pellprm\PGg$, at leading order (LO),
the photon, as illustrated in Fig.~\ref{fig:ppTo4LG}, can be emitted
from a quark line, including initial-state radiation, or
through the associated production of a Higgs boson and a \PZ boson ($\PZ\PH$),
with the Higgs boson decaying to $\PZ\PGg$,
which contributes about 3\%--6\% to the $\pptollllG$ yield, or
as final-state radiation (FSR) from a charged lepton.
Genuine $\ZZG$ intermediate states are produced from quark-line emission
and include a small $\PZ\PH$ contribution (first two diagrams of the same figure).

This Letter presents two measurements:
a search for triboson production $\Pp\Pp \to \ZZG$,
with both \PZ bosons decaying leptonically, and
an inclusive search for total $\pptollllG$,
including also FSR contributions.
The analysis uses final states with four charged leptons (electrons or muons)
and a photon, using data recorded by the CMS detector at $\sqrt{s} = 13\TeV$
during 2016--2018, corresponding to 138\fbinv of integrated luminosity.
Tabulated results are provided in the HEPData record for this analysis~\cite{hepdata}.

The CMS apparatus~\cite{CMS-CMS-00-001,CMS-PRF-21-001} is a multipurpose, nearly hermetic detector
designed to trigger on~\cite{CMS-TRG-17-001,CMS-TRG-12-001,CMS-TRG-19-001}
and identify~\cite{CMS-EGM-17-001,CMS-MUO-16-001,CMS-TRK-11-001}
electrons, muons, photons, and hadrons.
A global particle-flow algorithm~\cite{CMS-PRF-14-001} is used to reconstruct
individual particles in an event, combining information from
the all-silicon inner tracker, the crystal electromagnetic (ECAL) and
a brass and scintillator hadron (HCAL) calorimeters, operating inside a 3.8\unit{T}
superconducting solenoid, with data from the gas-ionization muon detectors
embedded in the flux-return yoke outside the solenoid.
The particles are used to reconstruct \PGt leptons, jets, and missing
transverse momentum~\cite{CMS-TAU-16-003,CMS-JME-13-004,CMS-JME-17-001}.

The $\pptollllG$ signal sample is simulated at next-to-leading order (NLO)
in quantum chromodynamics (QCD) and LO in EW interactions,
using the \MGvATNLO 2.6.2~\cite{MGatNLO,Frederix_2018} Monte Carlo (MC) generator.
The simulation of the $\PZ\PH$ signal
and of the $\qqZZll$ and \ttbar processes is performed at NLO QCD with
\POWHEG~v2.0~\cite{Nason:2004rx,Frixione:2007vw,Alioli:2010xd,Melia:2011tj,Nason:2013ydw},
and that of $\Pg\Pg \to \PZ\PZ \to 4\Pell$ at LO QCD with \MCFM~\cite{MCFM}.
The diboson production process $\PZ\PGg \to 2\Pell \PGg$
is simulated at NLO with \MGvATNLO, including off-shell effects.
Up to one additional parton is included in the matrix-element calculation.
Other rare background processes, such as $\PV\PV\PV$ ($\PV = \PZ, \PW$), $\ttbar\PW$, $\ttbar\PZ$,
$\ttbar\PW\PW$, $\ttbar\PZ\PZ$, $\PQt\PZ\PQq$, and $\PW\PZ\PGg$,
are simulated at NLO QCD with \MGvATNLO.

Hadronization and parton showering are simulated by interfacing the matrix-element generators
with \PYTHIA~8~\cite{Sjostrand:2015}, using the {CP5} tune~\cite{CMS-GEN-17-001}.
The interaction of particles with the CMS detector is modeled with \GEANTfour~\cite{GEANT}.
All samples are generated with the NNPDF~3.1 parton distribution
functions (PDFs)~\cite{NNPDF2015} at next-to-NLO in QCD.
The simulations include additional interactions in the same and neighboring bunch crossings, referred to as pileup.
The MC samples are reweighted so that the per-event true number of interactions
matches the average level of pileup observed in data for each year.

The signal sample $\pptollllG$
includes the triboson process $\Pp\Pp \to \PZoptst\PZoptst \PGg \to 4\Pell\PGg$,
and FSR contributions.
Their relative fractions have been studied at the generator level
by comparing two samples produced at LO with \MADGRAPH:
one forcing the presence of the intermediate $\ZZG$ state,
and a second where nonresonant and FSR diagrams are allowed.
Both the main signal $\pptollllG$
and the main background $\qqZZll$ samples
include FSR contributions.
In the latter, the photon is added by the parton shower simulation.
To avoid double counting, an event containing
a prompt photon that passes
the fiducial phase space definition of the inclusive $\pptollllG$ region
at the generator level
is removed from the main background sample.

The primary vertex is defined as the vertex corresponding to the hardest
scattering in the event, determined using tracking information alone~\cite{CMS-TDR-15-02}.
The kinematic variables of photons are obtained from ECAL energy deposits.
The four-momentum of electrons is determined from a combination of
the electron momentum at the primary interaction vertex, as measured by the tracker,
the energy of the corresponding ECAL cluster, and the summed energy of all
bremsstrahlung photons spatially compatible with originating from the electron.
The four-momentum of muons is derived from the curvature of the corresponding track.

The kinematic requirements for electrons and muons
reflect the detector acceptance and consist of
a threshold for the transverse momentum, $\pt^{\Pe(\PGm)} > 7\,(5)\GeV$, and
a requirement on the pseudorapidity, $\abs{\eta^{\Pe(\PGm)}} < 2.5\,(2.4)$.
Leptons must originate from the primary vertex, with a distance of closest approach
of less than 0.5\cm in the transverse plane and 1\cm along the beam axis.
The significance of the three-dimensional impact parameter of the track must satisfy
$\abs{\text{IP}}/{\sigma_\text{IP}} < 4$.
To ensure that leptons are isolated, the scalar sum of the transverse momenta
of the additional particles
within a cone of radius $\DR = \sqrt{\smash[b]{(\Delta\eta)^2+(\Delta\phi)^2}} = 0.3$
around the lepton momentum direction is required to be less than 35\% of $\pt^\Pell$.

Electrons are identified using a multivariate discriminant that exploits
characteristics of the electron's electromagnetic shower in the ECAL,
its matching to the associated track,
the presence of bremsstrahlung radiation,
observables that help discriminate against electrons from photon conversions,
and the isolation of the electron candidate~\cite{CMS-HIG-21-009}.
Muons are selected by applying loose requirements on the track quality
in the inner tracker and the muon system,
and on the compatibility of the associated calorimeter energy deposits
with the muon hypothesis.
The efficiency for low-\pt muons is improved
by including candidates whose inner track is matched
to segments in only one or two layers of the muon system.

For a better measurement of the \PZ boson mass, photons
emitted at small angles from leptons, $\DR(\PGg, \Pell) < 0.5$,
are recovered as follows.
Photons with $\abs{\eta^\PGg} < 2.4$, $\pt^\PGg > 2\GeV$,
and a relative isolation $I_\PGg < 1.8$,
computed as described in Ref.~\cite{CMS-HIG-21-009},
are considered.
Each photon is associated with the closest lepton in the event if
$\dRlGoverPtG < 0.012\GeV^{-2}$.
For each lepton, the photon with the lowest value of \dRlGoverPtG, if any, is selected.
The momentum of the photons identified by this algorithm
is added to that of the corresponding lepton.
These photons are excluded from the computation of the lepton isolation.

Signal photon candidates must be separated from any signal lepton by
$\DR(\PGg, \Pell) > 0.5$; this requirement increases the relative
contribution of the triboson component in the signal sample.
In addition, signal photons are required to have $\pt^\PGg > 20\GeV$,
$\abs{\eta^\PGg} < 2.4$ and $\abs{\eta^\PGg} \notin (1.444, 1.566)$,
corresponding to the ECAL barrel-endcap transition.
Electrons mimicking photons are rejected by requiring the absence of tracks
with a hit in the innermost layer of the pixel detector
that point to the photon cluster in the ECAL
and are matched to a reconstructed conversion vertex~\cite{CMS-EGM-17-001}.

A cutoff-based identifier and a multivariate discriminant~\cite{CMS-EGM-17-001}
are employed for the selection of signal photons.
The former uses the shape of the electromagnetic shower in the ECAL,
the energy deposited in the HCAL, and three observables related to the photon
candidate isolation.
A working point with an average efficiency of about~90\%
is used to identify ``tight'' photons.
``Loose'' photons have the same requirements, except
the shower shape and isolation from charged hadrons criteria are removed.
The multivariate discriminant employs 14 input variables,
including those used in the cutoff-based selection.
For this discriminant, only tight photons are defined,
using a working point with an average efficiency of about 80\%.

Corrections for the differences in reconstruction and selection efficiencies between
data and simulation are applied for muons~\cite{CMS-MUO-16-001,CMS-MUO-17-001},
electrons, and photons~\cite{CMS-EGM-17-001}.

Events are selected online using a combination of triggers~\cite{CMS-TRG-12-001}.
The most important ones require the presence of two loosely isolated leptons.
Additional triggers requiring three low-\pt leptons or a single electron or muon
are used to further increase the efficiency.
The overall trigger efficiency exceeds 98\%.

The event selection criteria are similar to those used in
recent Higgs boson measurements~\cite{CMS-HIG-21-009}.
Each event must contain two \PZ boson candidates,
each consisting of an oppositely charged, same-flavor pair of isolated leptons,
and a signal photon.
Both \PZ boson candidates are required to have a mass between 60 and 120\GeV.
The \PZ candidate with a mass closest to 91.2\GeV
is labeled $\PZ_1$, and the other $\PZ_2$.
The leading (subleading) lepton must satisfy $\pt > 20\,(10)\GeV$.

Each lepton must be separated from all others by $\DR(\Pell_i,\Pell_j) > 0.02$,
and electrons must be separated from muons by $\DR(\Pe,\PGm) > 0.05$.
Events containing pairs of oppositely charged leptons with an invariant mass
$m_{\Pellp\Pellprm} < 4\GeV$, computed without FSR photons, are rejected,
irrespective of lepton flavor.
Events containing four electrons or four muons are rejected
if the same four leptons can be used to form an alternative pair of $\PZ$ bosons,
$\PZ_\mathrm{a}$ and $\PZ_\mathrm{b}$, such that $m_{\PZ_\mathrm{b}} < 12\GeV$
and $m_{\PZ_\mathrm{a}}$ is closer to 91.2\GeV than the originally selected $\PZ_1$.
This requirement suppresses events with an on-shell \PZ boson and a low-mass resonance.
In the triboson signal region, an additional requirement to suppress FSR events is applied:
the minimum of the invariant masses of the two lepton pairs combined with the photon,
$\min(m_{\Pell\Pell\PGg}, m_{\Pellpr\Pellpr\PGg})$, must exceed 100\GeV.
This threshold was optimized through a dedicated generator-level study.

Decays of \PZ bosons to \PGt leptons, followed by subsequent decays to electrons
or muons, are strongly suppressed by the lepton \pt thresholds,
and their contribution amounts to less than 0.5\% of the total~$4\Pell\PGg$ yield.
If such events satisfy the selection criteria, they are treated as signal,
although they are not included at the generator level for the cross section determination.
Consequently, the correction for possible \PGt lepton decays is effectively accounted for
in the efficiency calculation.

For the inclusive $\pptollllG$ measurement, we present the results
using the cutoff-based photon selection.
This approach allows the definition of a control region (CR) with a sufficient number of events
for a background estimate, by requiring photons to pass a loose selection,
while failing the tight criterion.
In contrast, the limited number of events in the $\ZZGtollllG$ enriched region
precludes the use of CRs.
For this measurement, we present results obtained using the multivariate selection,
which provides stronger background rejection than the cutoff-based approach
and is better suited for the measurement of the very rare $\ZZGtollllG$ process.
In the rare case that multiple photons pass the selection,
the one with the highest transverse energy is retained.

The signal region (SR) is affected by backgrounds arising from the production of
nonprompt leptons and photons.
The CRs are defined to estimate these backgrounds
and to check the agreement between data and simulation.
Rare backgrounds are estimated from simulation.

Nonprompt leptons originate from processes in which at least one lepton in the
final state does not directly arise from the decay of a \PW or \PZ boson.
The main sources include decays of heavy-flavor mesons, which typically
produce nonisolated leptons, as well as light-flavor quark and gluon jets and electrons from
photon conversions.
This background is estimated using the method employed in several previous
analyses of four-lepton final states from $\PZ\PZ$ decays
(\eg, Refs.~\cite{CMS-SMP-16-001, CMS-SMP-17-006, CMS-SMP-20-001, CMS-SMP-22-001}),
and described in detail in Ref.~\cite{CMS-HIG-13-002}.
The nonprompt-lepton background yield in the SR is extrapolated
using data from two CRs defined with criteria similar to those of the SR,
except that one or both leptons of the $\PZ_2$ candidate fail the
tight isolation and identification requirements while passing looser criteria.
Events in these CRs are reweighted according to the lepton
misidentification rate measured in a dedicated CR containing exactly
one \PZ boson and one additional loose lepton ($\PZ+\Pell$)~\cite{CMS-HIG-21-009}.

\begin{figure}
\includegraphics[width=0.45\textwidth]{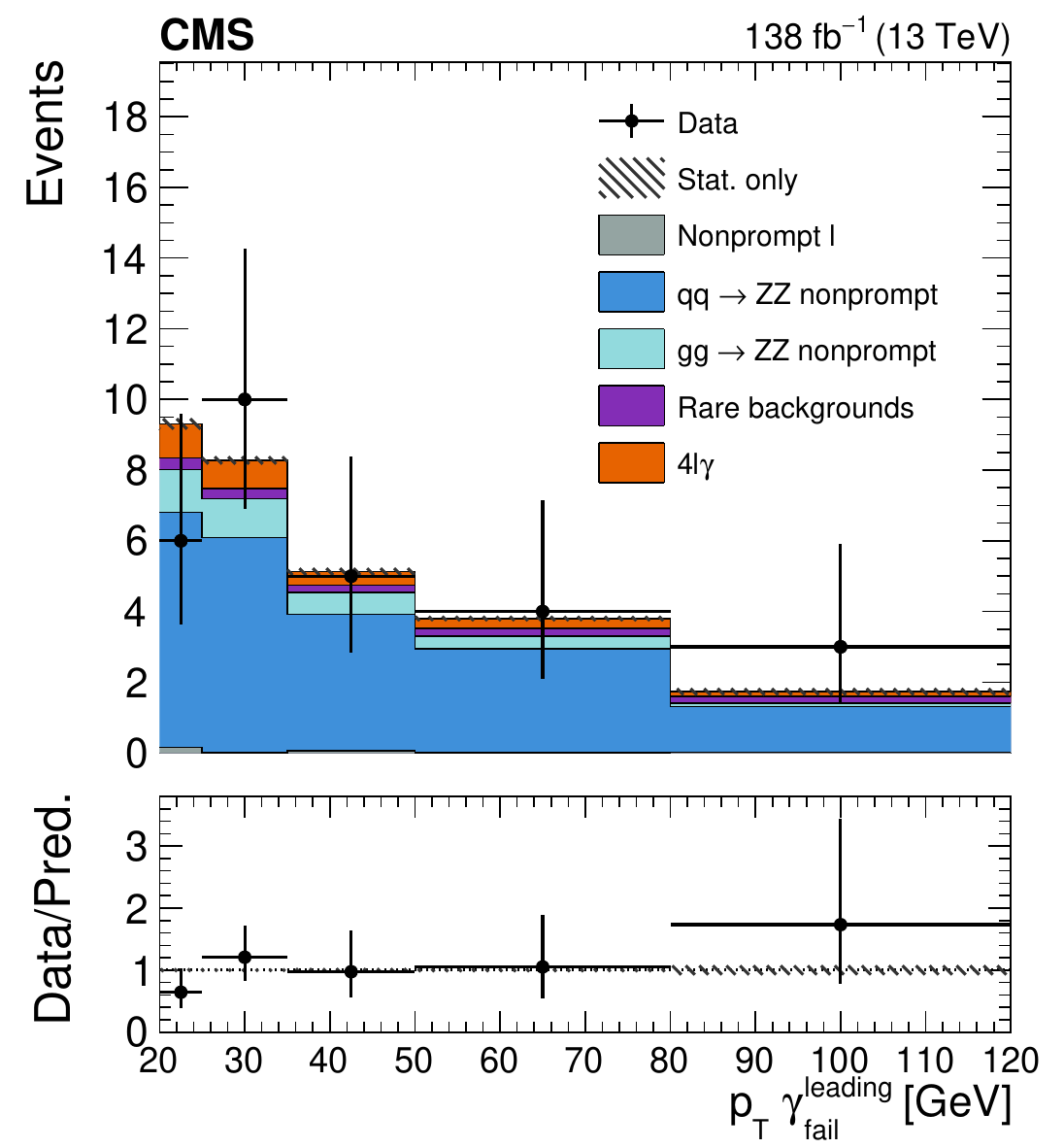}
\includegraphics[width=0.45\textwidth]{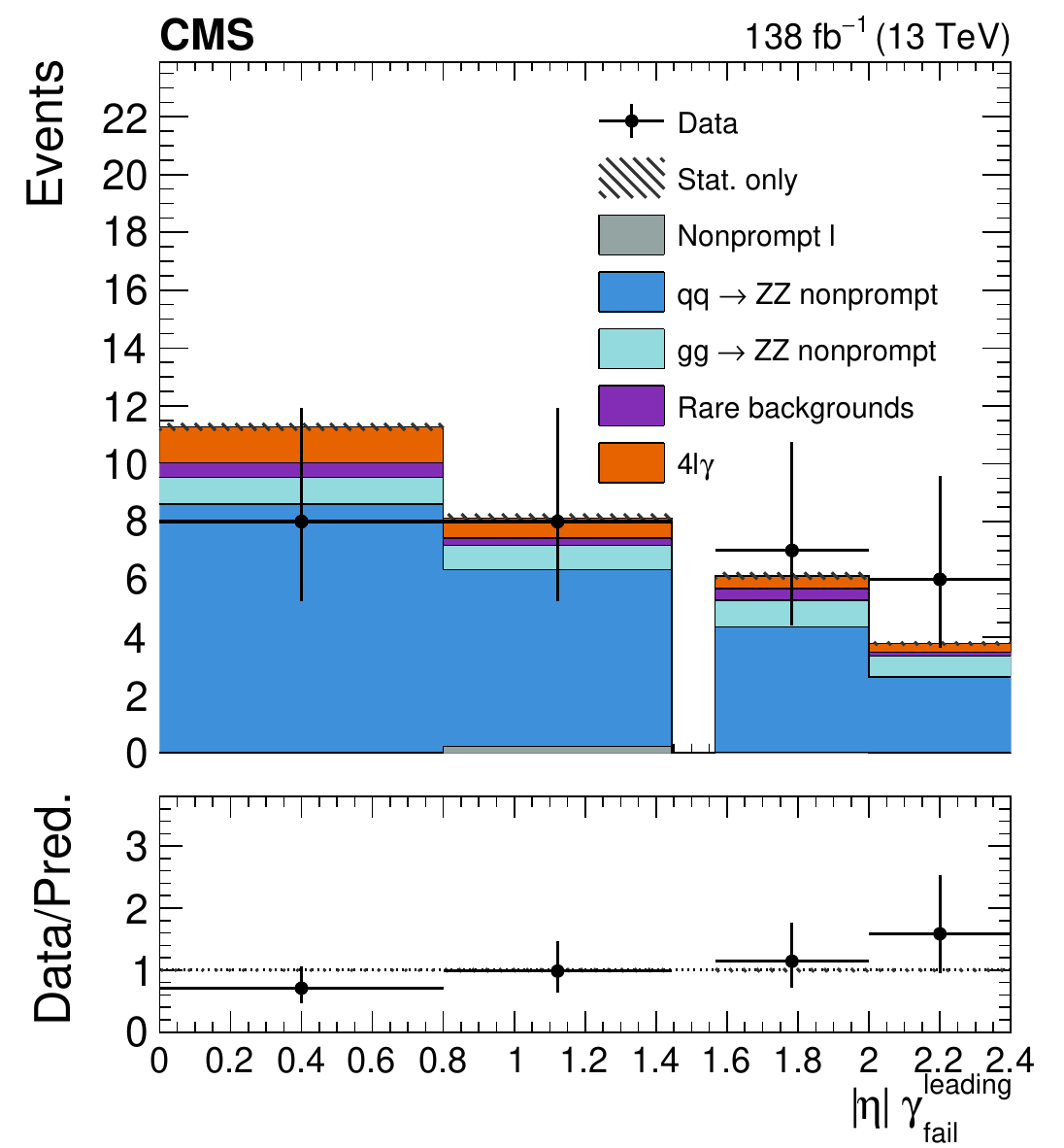}
\caption{
Transverse momentum (\cmsLeft) and pseudorapidity (\cmsRight) of the leading photon
passing the loose selection and failing the tight cutoff-based identification
in the nonprompt photon application region with four tight leptons.
The vertical bars represent the statistical uncertainties in the data.
The hatched area represents the statistical uncertainties in the simulation.
}
\label{fig:CR4P_1F}
\end{figure}

Nonprompt photons $(\phNP)$ arise mainly from jets---often containing genuine
photons from $\PGpz\to\PGg\PGg$ decays---depositing a large fraction of
their energy in the ECAL or from electrons
lacking an associated silicon tracker track and thus not rejected by the electron veto.
A data-driven tight-to-loose method is used to estimate this background.
The photon misidentification rate,
measured in bins of photon momentum and pseudorapidity,
in the $\PZ+\Pell$ CR after subtracting the $\PZ\PGg$ contributions from simulation,
is defined as the probability for a loose photon to satisfy the tight selection.
The application region is identical to the SR except that the photon fails
the tight selection.
Figure~\ref{fig:CR4P_1F} presents the $\pt^\PGg$ and $\abs{\eta^\PGg}$
distributions for data and simulation, showing agreement within uncertainties.
The pseudorapidity gap reflects the excluded barrel-endcap transition.
The nonprompt-photon yield is obtained by reweighting application region
events with the misidentification rate.
This approach, independent of the modeling of nonprompt-photon production,
provides the nonprompt-photon yields in the inclusive $\pptollllG$ SR.
In the triboson $\ZZG$ region this background is estimated from simulation.

The effects of experimental and theoretical systematic uncertainties for both
signal and background processes are
quantified by evaluating the impact of each source on the normalization and
shape of the relevant distributions of kinematic observables.

Theoretical uncertainties arising from the choice of the PDF set and
the value of the strong coupling constant, $\alpS$, are evaluated
according to the PDF4LHC prescription~\cite{Butterworth:2015oua}.
Their impacts are 0.9\% (1.4\%) and 0.3\% (1.3\%)
for the triboson (inclusive) measurement, respectively.
The impact of the QCD renormalization ($\mu_\mathrm{R}$) and factorization ($\mu_\mathrm{F}$) scales,
which account for missing higher-order contributions in perturbative calculations,
is evaluated from the envelope of the independent variations of
$\mu_\mathrm{R}$ and $\mu_\mathrm{F}$ by a factor of two around their central values,
while keeping $1/2 \leq \mu_\mathrm{R} / \mu_\mathrm{F} \leq 2$.
The resulting uncertainty is 1\% (5\%) for the triboson (inclusive) measurement.

The uncertainty in the integrated luminosity varies from 1.2\% to 2.5\% across data-taking periods,
with a total Run~2 uncertainty of 1.6\%~\cite{CMS-LUM-17-003, CMS-LUM-17-004, CMS-LUM-18-002}.
Its impact on the measured triboson (inclusive) cross section is 2.1\% (1.6)\%.

The estimates of the nonprompt-lepton and nonprompt-photon yields
are affected by uncertainties due to differences in background composition
between the measurement and application regions.
In the inclusive region the nonprompt background normalization uncertainties
are treated as a single uncertainty on the data-driven estimate of 50\%,
corresponding to a 3.9\% impact on the measurement.
In the triboson region, where simulation is used to estimate the nonprompt-photon
background and a data-driven method for the leptons, the uncertainty is 40\% for
leptons~\cite{CMS-SMP-20-001,CMS-SMP-22-001}, corresponding to a 1.7\% impact,
and 30\% are for photons, resulting in impacts of 4.5\% and 3.5\% from the
$\qqZZnonpro$ and $\ggZZnonpro$ yields respectively.

The ratio of simulated to observed nonprompt-photon yields from
$\PQq\PAQq/\Pg\Pg \to \PZ\PZ \phNP$ processes
is used to construct an alternative template for the background prediction
in the inclusive SR.
This shape uncertainty is treated as correlated across data-taking periods
and results in a 3.4\% uncertainty in the measurement.
Theoretical and experimental uncertainties affecting the prompt $\PZ\PGg$ component,
which is subtracted in the measurement region,
are propagated through the misidentification rate into the nonprompt-photon background estimate.
The combined impact on the signal strength uncertainty is approximately 4\%.

Efficiency scale factor uncertainties for electrons, muons, and photons affect both
the triboson and inclusive measurements by about 4--5\%, 1\%, and 2\%, respectively;
photon energy scale and resolution effects are negligible.
The offline \pt requirements for electrons and muons are more stringent
than the corresponding trigger thresholds~\cite{CMS-HIG-21-009,CMS-SMP-22-001}.
No trigger scale factor is applied, and a flat 2\% uncertainty is included.

Pileup modeling depends on the total inelastic $\Pp\Pp$ cross section~\cite{CMS-FSQ-15-005},
and its uncertainty is evaluated by varying this cross section by 5\%.

The signal definition for the inclusive process $\pptollllG$
at the generator-level mirrors the selection applied to reconstructed events,
except for the requirements
for both electrons and muons to have $\pt^\Pell > 5\GeV$ and $\abs{\eta^\Pell} < 2.5$,
and for the photon to be prompt and have $\abs{\eta^\PGg} < 2.4$.
At NLO, the predicted cross section in the fiducial region
$\sigma_\text{fid}(\pptollllG)$, with $\Pell = \Pe, \PGm$, is
$99.96 \pm 0.09 \unit{ab}$.
The fraction of triboson $\ZZG$ events in this region, estimated using the two
LO samples for the generator-level study, is approximately 38\%.
For the triboson signal definition, an additional requirement is imposed: the presence of a prompt generator-level
photon with $m_{\Pell\Pell\PGg} > 100\GeV$ for either of the two lepton pairs forming a \PZ candidate.
This condition reflects the selection applied at the reconstruction level.
The resulting fraction of triboson events in the signal sample $\pptollllG$
is 85\% and the predicted cross section, including the $\PZ\PH$ contribution, is
$47.56 \pm 0.04\unit{ab}$.

The cross section measurements are extracted using two independent maximum likelihood
template fits,
in the triboson and inclusive signal regions,
of the expected signal and background distributions
of the invariant mass of the four-lepton and photon system, $m_{4\Pell\PGg}$,
to the data for each data-taking period.
Each fit includes a single parameter of interest: the signal strength, defined as
the ratio of the measured fiducial cross section to the expected SM prediction.
This parameter quantifies the overall normalization of the signal relative to
the SM expectation;
nuisance parameters associated with systematic uncertainties are also included in the fit.
These results have been obtained using the CMS statistical analysis tool
\COMBINE~\cite{CMS-CAT-23-001}.

\begin{table}
  \centering
  \topcaption{
    Post-fit event yields in the triboson~$\ZZG$ (left) and inclusive~$4\Pell\PGg$ (right) regions.
    Uncertainties are statistical only.
    The row labeled \pptollllG~OOA (out-of-acceptance) gives the number
    of events in the $\pptollllG$ signal sample failing the fiducial selection.
  }
  \label{tab:yield_postfit}

  \begin{scotch}{l l c c}
    & & \makecell[c]{Triboson\\\ZZG} & \makecell[c]{Inclusive\\\pptollllG} \\
    \midrule
    \multicolumn{2}{c}{Signal} & $8.22 \pm 0.97$ & $21.4 \pm 1.5$ \\
    \hspace{-.5em}\multirow{3}{*}{$\left.
      \begin{tabular}{l}
        \qqZZ\\
        \ggZZ\\
        Nonprompt \Pell
      \end{tabular}\right\rbrace$\hspace{-.5em}}
    &\multirow{3}{*}{\makecell[l]{Background from\\nonprompt \Pell or \PGg}}
    & $0.77 \pm 0.07$ & \multirow{3}{*}{$2.11 \pm 0.76$}\\
    & & $0.68 \pm 0.04$ & \\
    & & $0.37 \pm 0.23$ & \\[\cmsTabSkip]
    \multicolumn{2}{c}{Rare backgrounds} & $0.37 \pm 0.05$ & $0.46 \pm 0.07$\\
    \multicolumn{2}{c}{\pptollllG~OOA} & $0.39 \pm 0.02$ & $0.24 \pm 0.02$\\[\cmsTabSkip]

    \multicolumn{2}{c}{Predicted number of events} & $10.8 \pm 1.0$ & $24.2 \pm 1.6$ \\
    \multicolumn{2}{c}{Observed number of events} & 11 & 24 \\
  \end{scotch}
\end{table}

\begin{figure}
  \centering
  \includegraphics[width=0.475\textwidth]{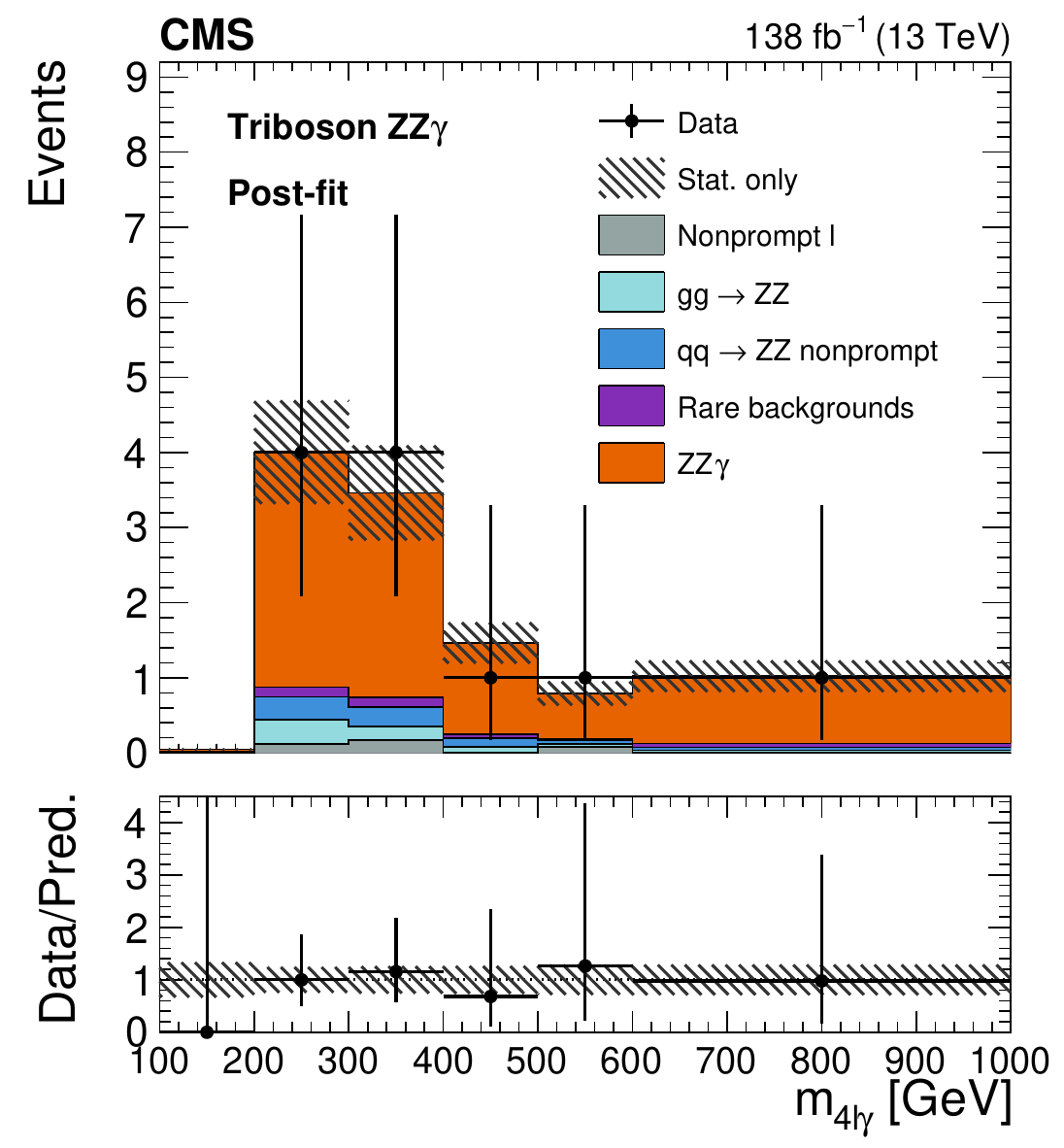}
  \includegraphics[width=0.475\textwidth]{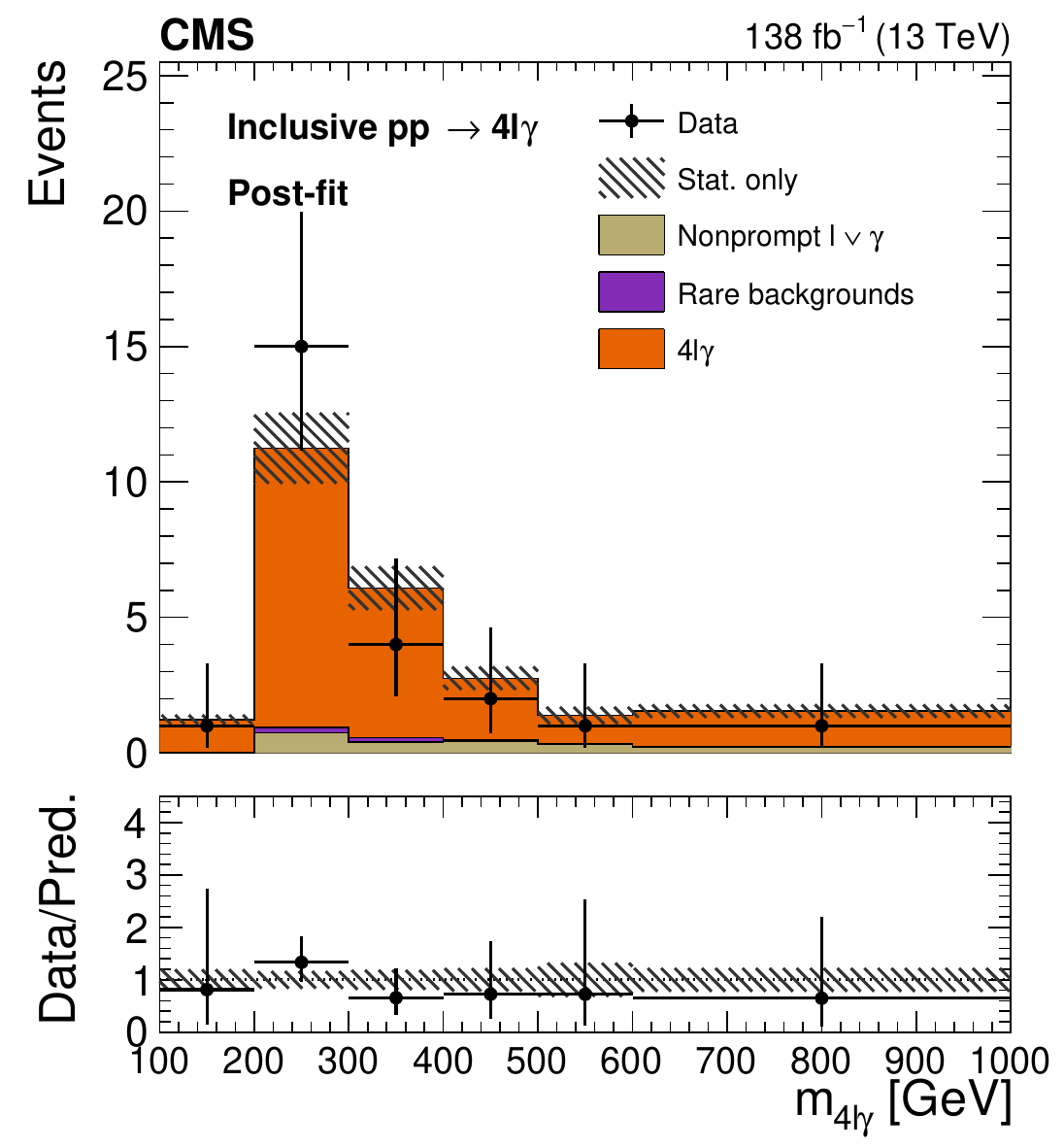}
  \caption{
    Post-fit mass distributions of the $\ZZG$ system in the triboson $\ZZG$ (\cmsLeft)
    and inclusive $\pptollllG$ (\cmsRight) regions,
    obtained with the cutoff-based and multivariate photon identification, respectively.
    The last bin includes the events exceeding the upper bound of the x-axis.
    The vertical bars represent the statistical uncertainties in the data.
    The hatched area represents the statistical uncertainties in the simulation.
  }
  \label{fig:dataMC_Run2_SR4P_1P}
\end{figure}

The data yields and fitted numbers of signal and background events
in the two SRs are given in Table~\ref{tab:yield_postfit}.
The nonprompt-photon background, which constitutes
more than 75\% of the total background, represents
8.3\% of the post-fit yield in the inclusive $\pptollllG$ SR.
Events in the signal sample that pass the selection but fail the signal definition
(out-of-acceptance, OOA) are treated as an additional background contribution.
Figure~\ref{fig:dataMC_Run2_SR4P_1P} shows the post-fit distributions
of $m_{4\Pell\PGg}$ in the triboson and inclusive SRs.

In the triboson SR, the statistical uncertainty dominates,
accounting for over 98\% of the total uncertainty.
The systematic uncertainties with the largest impact arise from
the normalization of the $\qqZZnonpro$ background and the electron efficiency.
The fit yields a signal strength of
$\mu =
1.25^{+0.56}_{-0.45}\stat\allowbreak {}^{+0.01}_{-0.01}\thy {}^{+0.03}_{-0.02}\lum {}^{+0.10}_{-0.08}\syst
=
1.25^{+0.57}_{-0.46},$
corresponding to a fiducial cross section of
$\sigma_{\text{fid}}(\pptoZZGtollllG) =
60^{+26}_{-22}\stat {}^{+1}_{-1}\thy {}^{+2}_{-1}\lum\allowbreak {}^{+5}_{-4}\syst
\unit{ab}
= 60^{+27}_{-22}\unit{ab}$.
The observed (expected) significance of $\ZZG$ triboson production
with respect to the background-only hypothesis is
3.7\,(3.1) standard deviations.
{\tolerance=800
In the inclusive $\pptollllG$ region, the statistical uncertainty also dominates,
accounting for over 90\% of the total.
The leading systematic uncertainties arise from the
renormalization and factorization scales, the electron efficiency,
and the shape and normalization of the nonprompt-photon background.
The fit yields a signal strength of
$\mu =
1.56^{+0.38}_{-0.33}\stat \allowbreak^{+0.04}_{-0.06}\thy \allowbreak^{+0.03}_{-0.02}\lum \allowbreak^{+0.09}_{-0.08}\syst
=
1.56^{+0.40}_{-0.35},$
corresponding to a fiducial cross section of
$\sigma_{\text{fid}}(\pptollllG) =
156^{+38}_{-33}\stat \allowbreak^{+4}_{-5}\thy \allowbreak^{+3}_{-2}\lum \allowbreak^{+9}_{-8}\syst
\unit{ab}
=
156^{+39}_{-35}
\unit{ab}.$
The observed (expected) significance of the signal relative to the background-only hypothesis is
$5.0$ ($4.2$) standard deviations.
\par}

In summary, this Letter presents the first evidence in CMS for triboson production,
$\Pp\Pp \to \ZZG$, where both \PZ bosons decay leptonically to electrons or muons,
in proton-proton collisions at $\sqrt{s}=13\TeV$, collected in 2016--2018
and corresponding to an integrated luminosity of $138\fbinv$.
The fiducial cross section,
$\sigma_{\text{fid}} = 60^{+27}_{-22}\unit{ab}$, is the smallest ever measured at the LHC.
In a more inclusive signal region with FSR contributions, $\pptollllG$ production is observed for the first time,
with a fiducial cross section of $156^{+39}_{-35}\unit{ab}$.
The results are in good agreement with standard model predictions.

\begin{acknowledgments}
We congratulate our colleagues in the CERN accelerator departments for the excellent performance of the LHC and thank the technical and administrative staffs at CERN and at other CMS institutes for their contributions to the success of the CMS effort. In addition, we gratefully acknowledge the computing centers and personnel of the Worldwide LHC Computing Grid and other centers for delivering so effectively the computing infrastructure essential to our analyses. Finally, we acknowledge the enduring support for the construction and operation of the LHC, the CMS detector, and the supporting computing infrastructure provided by the following funding agencies: SC (Armenia), BMBWF and FWF (Austria); FNRS and FWO (Belgium); CNPq, CAPES, FAPERJ, FAPERGS, and FAPESP (Brazil); MES and BNSF (Bulgaria); CERN; CAS, MoST, and NSFC (China); MINCIENCIAS (Colombia); MSES and CSF (Croatia); RIF (Cyprus); SENESCYT (Ecuador); ERC PRG and PSG, TARISTU24-TK10 and MoER TK202 (Estonia); Academy of Finland, MEC, and HIP (Finland); CEA and CNRS/IN2P3 (France); SRNSF (Georgia); BMFTR, DFG, and HGF (Germany); GSRI (Greece); NKFIH (Hungary); DAE and DST (India); IPM (Iran); SFI (Ireland); INFN (Italy); MSIT and NRF (Republic of Korea); MES (Latvia); LMTLT (Lithuania); MOE and UM (Malaysia); BUAP, CINVESTAV, CONACYT, LNS, SEP, and UASLP-FAI (Mexico); MOS (Montenegro); MBIE (New Zealand); PAEC (Pakistan); MSHE, NSC, and NAWA (Poland); FCT (Portugal); MESTD (Serbia); MICIU/AEI and PCTI (Spain); MOSTR (Sri Lanka); Swiss Funding Agencies (Switzerland); MST (Taipei); MHESI (Thailand); TUBITAK and TENMAK (T\"{u}rkiye); NASU (Ukraine); STFC (United Kingdom); DOE and NSF (USA).
\end{acknowledgments}

\bibliography{auto_generated}
\cleardoublepage \appendix\section{The CMS Collaboration \label{app:collab}}\begin{sloppypar}\hyphenpenalty=5000\widowpenalty=500\clubpenalty=5000\cmsinstitute{Yerevan Physics Institute, Yerevan, Armenia}
{\tolerance=6000
A.~Hayrapetyan, V.~Makarenko\cmsorcid{0000-0002-8406-8605}, A.~Tumasyan\cmsAuthorMark{1}\cmsorcid{0009-0000-0684-6742}
\par}
\cmsinstitute{Institut f\"{u}r Hochenergiephysik, Vienna, Austria}
{\tolerance=6000
W.~Adam\cmsorcid{0000-0001-9099-4341}, L.~Benato\cmsorcid{0000-0001-5135-7489}, T.~Bergauer\cmsorcid{0000-0002-5786-0293}, M.~Dragicevic\cmsorcid{0000-0003-1967-6783}, P.S.~Hussain\cmsorcid{0000-0002-4825-5278}, M.~Jeitler\cmsAuthorMark{2}\cmsorcid{0000-0002-5141-9560}, N.~Krammer\cmsorcid{0000-0002-0548-0985}, A.~Li\cmsorcid{0000-0002-4547-116X}, D.~Liko\cmsorcid{0000-0002-3380-473X}, M.~Matthewman, J.~Schieck\cmsAuthorMark{2}\cmsorcid{0000-0002-1058-8093}, R.~Sch\"{o}fbeck\cmsAuthorMark{2}\cmsorcid{0000-0002-2332-8784}, M.~Shooshtari\cmsorcid{0009-0004-8882-4887}, M.~Sonawane\cmsorcid{0000-0003-0510-7010}, N.~Van~Den~Bossche\cmsorcid{0000-0003-2973-4991}, W.~Waltenberger\cmsorcid{0000-0002-6215-7228}, C.-E.~Wulz\cmsAuthorMark{2}\cmsorcid{0000-0001-9226-5812}
% ...
\par}
\vskip\cmsinstskip
\dag:~Deceased\\
$^{1}$Also at Yerevan State University, Yerevan, Armenia\\
$^{2}$Also at TU Wien, Vienna, Austria\\
% ...
\end{sloppypar}
\end{document}